\documentclass[journal, final, letterpaper, twocolumn, twoside]{IEEEtran}

\usepackage{cite} 

\usepackage[cmex10]{amsmath}
\usepackage{amssymb, amsthm, eucal}
\usepackage{array}

\usepackage{graphicx, epsfig}
\usepackage[tight, footnotesize]{subfigure}

\newtheorem{proposition}{Proposition}
\newtheorem{example}{Example}
\newtheorem{assumption}{Assumption}

\def\e{\epsilon}
\def\g{\gamma}
\def\re{\Re}
\def\rn{\Re^n}
\def\m{\mu}
\def\d{\delta}

\def\pttj[#1]{\frac{\partial #1}{\partial \theta_j}}
\def\tpttj[#1]{\tfrac{\partial #1}{\partial \theta_j}}
\def\tpttk[#1]{\tfrac{\partial #1}{\partial \theta_k}}
\def\tpttjk[#1]{\tfrac{\partial^2 #1}{\partial \theta_j \partial \theta_k}}

\def \l {\lambda}

\def\p{\pi}

\def\argmin{\mathop {\arg \min}}

\def\smskip{\smallskip}
\def\pn{\par\noindent}
\def\bl{\big}
\def\br{\big}

\def\tl{\tilde}
\def\lf{\left}
\def\ri{\right}

\def\old#1{}

\def\eqref#1{(\ref{#1})}
\def\proof{\pn {\it Proof\/}:\ }

\begin{document}

\title{Value and Policy Iteration in Optimal Control and Adaptive Dynamic Programming}

\old{
\author{\IEEEauthorblockN{Dimitri P. Bertsekas}
\IEEEauthorblockA{
Massachusetts Institute of Technology}
}
}

\author{Dimitri P. Bertsekas
\thanks{Dimitri Bertsekas is with the Dept.\ of Electr.\ Engineering and
Comp.\ Science, M.I.T., Cambridge, Mass., 02139.
        {\tt\small dimitrib@mit.edu}}%
\thanks{Many thanks are due to Huizhen (Janey) Yu for collaboration and many helpful discussions in the course of related works.}%
}

 \maketitle

\begin{abstract}
In this paper, we consider discrete-time infinite horizon problems of optimal control to a terminal set of states. These are the problems  that are often taken as the starting point for adaptive dynamic programming. Under very general assumptions, we establish the uniqueness of solution of Bellman's equation, and we provide  convergence results for value and policy iteration.
\end{abstract}

{
\setlength{\unitlength}{1cm}
\begin{picture}(0,0)(0,0)
\put(-0.4,7.1){\bf Report LIDS-P-3174, May 2015 (Revised Sept.\ 2015)\ \ \ \ \ To appear in IEEE Transactions on Neural Networks, 2015
 }
\end{picture}
}

\section{Introduction} \label{sec-intro}

\pn 
In this paper we consider a deterministic discrete-time optimal control problem involving the system
\begin{equation}\label{docsys}
x_{k+1}=f(x_k,u_k),\qquad k=0,1,\ldots,
\end{equation}
where $x_k$ and $u_k$ are the state and control at stage $k$, lying in sets $X$ and $U$, respectively, and $f$ is a function mapping $X\times U$ to $X$. The control $u_k$ must be chosen from a constraint set $U(x_k)\subset U$ that may depend on the current state $x_k$. The cost for the $k$th stage, denoted $g(x_k,u_k)$, is assumed nonnnegative and may possibly take the value $\infty$:
\begin{equation}\label{nonnegcost}
0\le g(x_k,u_k)\le\infty,\qquad x_k\in X,\ u_k\in U(x_k),
\end{equation}
[values $g(x_k,u_k)=\infty$ may be used to model constraints on $x_k$, for example]. We are interested in  feedback policies of the form $\p=\{\m_0,\m_1,\ldots\}$, where each $\m_k$ is a function mapping every $x\in X$ into the control $\m_k(x)\in U(x)$. The set of all  policies is denoted by $\Pi$. Policies of the form $\p=\{\m,\m,\ldots\}$ are called {\it stationary\/}, and for convenience, when confusion cannot arise, will be denoted by $\m$. No restrictions are placed on $X$ and $U$: for example, they may be finite sets as in classical shortest path problems involving a graph, or they may be continuous spaces as in classical problems of control to the origin or some other terminal set.

Given an initial state $x_0$, a policy $\p=\{\m_0,\m_1,\ldots\}$ when applied to the system \eqref{docsys}, generates a unique sequence of state control pairs $\big(x_k,\m_k(x_k)\big)$, $k=0,1,\ldots,$ with cost 
\begin{equation}\label{xdefcostjpdet}
J_\p(x_0)= \lim_{k\to\infty} \sum_{t=0}^{k}
g\bl(x_t,\mu_t(x_t)\br),\qquad x_0\in X,
\end{equation}
[the limit exists thanks to the nonnegativity assumption \eqref{nonnegcost}]. We view $J_\p$ as a function over $X$ that takes values in $[0,\infty]$. We refer to it as the cost function of $\p$. For a stationary policy $\m$, the corresponding cost function is denoted by $J_\m$.
The optimal cost function is defined as
\begin{equation*}
 J^*(x)=\inf_{\p\in\Pi}J_\p(x),\qquad x\in X,
\end{equation*}
and a policy $\p^*$ is said to be optimal if it attains the minimum of $J_\p(x)$ for all $x\in X$, i.e.,
\begin{equation*}
J_{\p^*}(x)=\inf_{\p\in\Pi}J_\p(x)= J^*(x),\qquad \forall\ x\in X.
\end{equation*}

In the context of dynamic programming (DP for short), one hopes to prove that the optimal cost function $ J^*$ satisfies Bellman's equation:
\begin{equation} \label{bellmaneq}
 J^*(x)=\inf_{u\in U(x)}\big\{g(x,u)+ J^*\big(f(x,u)\big)\big\},\qquad \forall\ x\in X,
\end{equation}
and that an optimal stationary policy may be obtained through the minimization in the right side of this equation. Note that Bellman's equation generically has multiple solutions, since adding a positive constant to any solution produces another solution. A classical result, stated in Prop.\ 4(a) of Section II, is that the optimal cost function $J^*$ is the ``smallest" solution of Bellman's equation. In this paper we will focus on deriving conditions under which $J^*$ is the unique solution within a certain restricted class of functions.

In this paper, we will also consider finding $J^*$ with the classical algorithms of value iteration (VI for short) and policy iteration (PI for short). The VI algorithm starts from some nonnegative function $J_0:X\mapsto [0,\infty]$, and generates a sequence of functions $\{J_k\}$ according to 
\begin{equation} \label{vieq}
J_{k+1}=\inf_{u\in U(x)}\big\{g(x,u)+J_k\big(f(x,u)\big)\big\}.
\end{equation}
We will derive conditions under which $J_k$ converges to $J^*$ pointwise.

The PI algorithm starts from a stationary policy $\m^0$, and generates a sequence of stationary policies $\{\m^k\}$ via a sequence of policy evaluations to obtain $J_{\m^k}$ from the equation
\begin{equation} \label{poleval}
J_{\m^k}(x)=g\big(x,\m^k(x)\big)+J_{\m^k}\big(f\big(x,\m^k(x)\big)\big),\qquad x\in X,
\end{equation}
interleaved with policy improvements to obtain $\m^{k+1}$ from $J_{\m^k}$ according to
\begin{equation} \label{polimprove}
\m^{k+1}(x)\in \argmin_{u\in U(x)}\big\{g(x,u)+J_{\m^k}\big(f(x,u)\big)\big\},\qquad x\in X.
\end{equation}
We implicitly assume here is that $J_{\m^k}$ satisfies Eq.\ \eqref{poleval}, which is true under the cost nonnegativity assumption \eqref{nonnegcost} (cf.\ Prop.\ 4 in the next section). Also  for the PI algorithm to be well-defined, the minimum in Eq.\ \eqref{polimprove} should be attained for each $x\in X$, which is true under some  conditions that guarantee compactness of the level sets 
\begin{equation*}
\big\{u\in U(x)\mid g(x,u)+J_{\m^k}\big(f(x,u)\big)\le \l\big\},\qquad \l\in\re.
\end{equation*}
We will derive conditions under which $J_{\m^k}$ converges to $J^*$ pointwise.

In this paper, we will address the preceding questions, for the case where there is a nonempty stopping set $X_s\subset X$, which consists of cost-free and absorbing states in the sense that
\begin{equation} \label{absorbe}
g(x,u)=0,\qquad x=f(x,u),\qquad \forall\ x\in X_s,\ u\in U(x).
\end{equation}
Clearly, $ J^*(x)=0$ for all $x\in X_s$, so  the set $X_s$ may be viewed as a desirable set of termination states  that we are trying to reach or approach with minimum total cost. We will assume in addition that $ J^*(x)>0$ for $x\notin X_s$, so that
\begin{equation} \label{stopset}
X_s=\big\{x\in X\mid  J^*(x)=0\big\}.
\end{equation}
In the applications of primary interest, $g$ is usually taken to be strictly positive outside of $X_s$ to encourage asymptotic convergence of the generated state sequence to $X_s$, so this assumption is natural and often easily verifiable.
Besides $X_s$, another interesting subset of $X$ is 
\begin{equation*}
X_f=\big\{x\in X\mid  J^*(x)<\infty\big\}.
\end{equation*}
Ordinarily, in practical applications, the states in $X_f$ are those from which one can reach the stopping set $X_s$, at least asymptotically.

For an initial state $x$, we say that a policy $\p$ {\it terminates} starting from $x$ if the state sequence $\{x_k\}$ generated starting from $x$ and using $\p$ reaches $X_s$ in finite time, i.e., satisfies $x_{\bar k}\in X_s$ for some index $\bar k$. A key assumption in this paper is that the optimal cost $ J^*(x)$ (if it is finite) can be approximated arbitrarily closely by using policies that terminate from $x$. In particular, in all the results and discussions of the paper we make the following assumption (except for Prop.\ 5, which provides conditions under which the assumption holds).

\begin{assumption} \label{assumptiondetoc}
The cost nonnegativity condition \eqref{nonnegcost} and stopping set conditions \eqref{absorbe}-\eqref{stopset} hold. Moreover, for every pair $(x,\e)$ with $x\in X_f$ and $\e>0$, there exists a policy $\p$ that terminates starting from $x$ and satisfies $J_\pi(x)\le J^*(x)+\e$.
  \end{assumption}
 
Specific and easily verifiable conditions that imply this assumption will be given in Section IV.  A prominent case is when $X$ and $U$ are finite, so the problem becomes a deterministic shortest path problem with nonnegative arc lengths.  If all cycles of the state transition graph have positive length, all policies $\p$ that do not terminate from a state $x\in X_f$ must satisfy $J_\p(x)=\infty$, implying that there exists an optimal policy that terminates from all $x\in X_f$. Thus, in this case Assumption \ref{assumptiondetoc}
 is naturally satisfied.

When $X$ is the $n$-dimensional Euclidean space $\rn$, a primary case of interest for this paper, it may easily happen that the optimal policies are not terminating from some $x\in X_f$, but instead the optimal state trajectories may approach $X_s$ asymptotically. This is true for example in the classical linear-quadratic optimal control problem, where $X=\rn$, $X_s=\{0\}$, $U=\re^m$, $g$ is positive semidefinite quadratic, and $f$ represents a linear system of the form $x_{k+1}=Ax_k+Bu_k$, where $A$ and $B$ are given matrices. However, we will show in Section IV that Assumption \ref{assumptiondetoc} is satisfied under some natural and easily verifiable conditions. 

\old{
For example, it is satisfied assuming that $g$ is strictly positive outside $X_s$ in the sense that for each $\d>0$ there exists $\e>0$ such that 
\begin{equation*}
\inf_{u\in U(x)}g(x,u)\ge\e,\qquad \forall\ x\in X\hbox{ such that } \hbox{dist}(x,X_s)\ge \d,
\end{equation*}
where for all $x\in X$,  $\hbox{dist}(x,X_s)$ denotes the minimum distance from $x$ to $X_s$,
\begin{equation*}
\hbox{dist}(x,X_s)=\inf_{y\in X_s}\|x-y\|,\qquad x\in X,
\end{equation*}
and also that for every $\e>0$, there exists a $\d_\e>0$ such that for each $x\in  X_f$ with 
$\hbox{dist}(x,X_s)\le \d_\e,$
there is a policy $\pi$ that terminates from $x$ and satisfies $J_\p(x)\le \e$.
The latter condition  is a ``local controllability" assumption implying that the state can be steered into $X_s$ with arbitrarily small cost from any starting state that is sufficiently close to $X_s$, and can be easily checked in many applications. 
}

Regarding notation, we denote by $\re$ and $\rn$ the real line and $n$-dimensional Euclidean space, respectively. We denote by $E^+(X)$ the set of all functions $J:X\mapsto[0,\infty]$, and by ${\cal J}$ the set of functions
\begin{equation} \label{regsets}
{\cal J}=\big\{J\in E^+(X)\mid J(x)=0,\,\forall\ x\in X_s\big\}.
\end{equation}
Since $X_s$ consists of cost-free and absorbing states [cf.\  Eq.\ (\ref{absorbe})], the set ${\cal J}$ contains the cost function $J_\p$ of all policies $\p$, as well as $ J^*$. In our terminology,  all equations, inequalities, and convergence limits involving functions are meant to be pointwise. Our main results are given in the following three propositions. 

\begin{proposition}[Uniqueness of Solution of Bellman's Equation] \label{propdetoc1}
Let Assumption \ref{assumptiondetoc} hold. The optimal cost function 
$ J^*$ is the unique solution of Bellman's equation  (\ref{bellmaneq}) within the set of  functions ${\cal J}$.\end{proposition}

There are well-known examples where $g\ge0$ but Assumption \ref{assumptiondetoc} does not hold, and there are additional solutions  of Bellman's equation within $ {\cal J}$. The following is a two-state shortest path example, which is discussed in more detail in [12], Section 3.1.2, and [14], Example 1.1.

\begin{example}[Counterexample for Uniqueness of Solution of Bellman's Equation] \label{example1}
Let $X=\{0,1\}$, where 0 is the unique cost-free and absorbing state, $X_s=\{0\}$, and assume that at state 1 we can stay at 1 at no cost, or move to 0 at cost 1. Here $J^*(0)=J^*(1)=0$, so Eq.\ (\ref{stopset}) is violated. It can be seen that 
$${\cal J}=\big\{J\mid J^*(0)=0,\ J^*(1)\ge 0\big\},$$
and that Bellman's equation is
\begin{equation*}
J^*(0)=J^*(0),\qquad J^*(1)=\min\big\{J^*(1),\,1+J^*(0)\big\}.
\end{equation*}
It can be seen that Bellman's equation has infinitely many solutions within ${\cal J}$, the set $\big\{J\mid J(0)=0,\, 0\le J(1)\le 1\big\}$.
\end{example}

\begin{proposition}[Convergence of VI] \label{propdetvi}
Let Assumption \ref{assumptiondetoc} hold.
\begin{itemize}
\item[]
\item [(a)]  The VI sequence $\{J_k\}$ generated by Eq.\ (\ref{vieq}) converges pointwise to $ J^*$ starting from any function $J_0\in{\cal J}$ with $J_0\ge  J^*$.
\item [(b)] Assume further that $U$ is a metric space, and the sets $U_k(x,\l)$ given by
$$U_k(x,\l)=\big\{u\in U(x)\mid g(x,u)+J_k\big(f(x,u)\big)\le \l\big\},$$
 are compact for all $x\in X$, $\l\in \re$, and $k$, where $\{J_k\}$ is the VI sequence $\{J_k\}$ generated by Eq.\ (\ref{vieq}) starting from $J_0\equiv0$. Then the VI sequence $\{J_k\}$ generated by Eq.\ (\ref{vieq}) converges pointwise to $ J^*$ starting from any function $J_0\in {\cal J}$.
\end{itemize}
 \end{proposition}

The compactness assumption of Prop.\ \ref{propdetvi}(b) is satisfied if $U(x)$ is finite for all $x\in X$. Other easily verifiable assumptions implying this compactness assumption will be given later. Note that when there are  solutions to Bellman's equation within ${\cal J}$,  in addition to $ J^*$, VI will not converge to $ J^*$ starting from any of these solutions. However, it is also possible that Bellman's equation has $ J^*$ as its unique solution within ${\cal J}$, and yet VI does not converge to $ J^*$ starting from the zero function because the compactness assumption of Prop.\ \ref{propdetvi}(b) is violated. There are several examples of this type in the literature, and the following example, an adaptation of Example 4.3.3 of [12], is a deterministic problem for which Assumption \ref{assumptiondetoc} is satisfied.

\begin{example}[Counterexample for Convergence of VI] \label{example2}Let $X=[0,\infty)\cup\{s\}$, with $s$ being a cost-free and absorbing state, and let $U=(0,\infty)\cup\{\bar u\}$, where $\bar u$ is a special stopping control, which moves the system from states $x\ge0$ to state $s$ at unit cost. The system has the form
$$
x_{k+1}=\begin{cases}
x_k+u_k& \text{if $x_k\ge0$ and $u_k\ne \bar u$,}\\
s& \text{if $x_k\ge0$ and $u_k=\bar u$,}\\
s& \text{if $x_k=s$ and $u_k\in U$.}
\end{cases}
$$
The cost per stage has the form
$$
g(x_k,u_k)=\begin{cases}
x_k& \text{if $x_k\ge0$ and $u_k\ne \bar u$,}\\
1& \text{if $x_k\ge0$ and $u_k=\bar u$,}\\
0& \text{if $x_k=s$ and $u_k\in U$.}
\end{cases}
$$
Let also $X_s=\{s\}$. Then it can be verified that
$$
J^*(x)=\begin{cases}
1& \text{if $x\ge0$,}\\
0& \text{if $x=s$,}
\end{cases}
$$
and that an optimal policy is to use the  stopping control $\bar u$ at every state (since using any other control at states $x\ge0$, leads to unbounded accumulation of positive cost). Thus it can be seen that Assumption \ref{assumptiondetoc} is satisfied. On the other hand, the VI algorithm is
$$J_{k+1}(x)=
\min\left\{1+J_k(s),\,\inf_{u\ge 0}\big\{x+J_k(x+u)\big\}\right\}$$
for  $x\ge0$, and 
$J_{k+1}(s)=J_k(s)$,
and it can be verified by induction that starting from $J_0\equiv0$, the sequence $\{J_k\}$ is given for all $k$ by
$$
J_k(x)=\begin{cases}
\min\{1,\,k x\}& \text{if $x\ge0$,}\\
0& \text{if $x=s$.}
\end{cases}
$$
Thus $J_k(0)=0$ for all $k$, while $J^*(0)=1$, so the VI algorithm fails to converge for the state $x=0$. The difficulty here is that the compactness assumption of Prop.\ \ref{propdetvi}(b) is violated.
\end{example}

\vskip-1.0pc

\begin{proposition}[Convergence of PI] \label{propdetpolicyit}
Let Assumption \ref{assumptiondetoc} hold. A sequence $\{J_{\m^k}\}$ generated by the PI algorithm (\ref{poleval}), (\ref{polimprove}), satisfies $J_{\m^k}(x)\downarrow  J^*(x)$ for all $x\in X$.
\end{proposition}

It is implicitly assumed in the preceding proposition that the PI algorithm is well-defined in the sense that the minimization in the policy improvement operation (\ref{polimprove}) can be carried out for every $x\in X$. Easily verifiable conditions that guarantee this also guarantee the compactness condition of Prop.\ \ref{propdetvi}(b), and will be noted following Prop.\ 4 in the next section. Moreover, in Section IV we will prove a similar convergence result for a variant of the PI algorithm where the policy evaluation is carried out approximately through a finite number of VIs. 

\begin{example}[Counterexample for Convergence of PI] \label{example3}For a simple  example where the PI sequence $J_{\m^k}$ does not converge to $ J^*$ if Assumption \ref{assumptiondetoc} is violated, consider the two-state shortest path Example \ref{example2}. 
Let $\m$ be the suboptimal policy that moves from state 1 to state 0. Then $J_\m(0)=0$, $J_\m(1)=1$, and it can be seen that $\m$ satisfies the policy improvement equation
$$\m(1)\in\arg\min\big\{1+J_\m(0),\, J_\m(1)\big\}.$$
Thus PI may stop with the suboptimal policy $\m$.
\end{example}

The results of the preceding three propositions are new at the level of generality given here. For example there has been no proposal of a valid PI algorithm in the classical literature on nonnegative cost infinite horizon Markovian decision problems (exceptions are special cases such as linear-quadratic problems [23]). The ideas of the present paper stem from a more general analysis regarding the convergence of VI, which was presented recently in the author's research monograph on abstract DP [Ber12], and various extensions given in the recent papers [13], [14]. Two more papers of the author, coauthored with H.\ Yu, deal with issues that relate in part to the intricacies of the convergence of VI and PI in undiscounted infinite horizon DP  [35], [5].

The paper is organized as follows. In Section II we provide background and references, which place in context our results and methods of analysis in relation to the literature. In Section III we give the proofs of Props.\ \ref{propdetoc1}-\ref{propdetpolicyit}. In Section IV we discuss special cases and easily verifiable conditions that imply our assumptions, and we provide extensions of our analysis.

\vskip-1.2pc

\section{Background}

\pn The issues discussed in this paper have received attention since the 60's, originally in the work of Blackwell [15], who considered the case $g\le0$, and the work by Strauch (Blackwell's PhD student) [30], who considered the case $g\ge0$. For textbook accounts we refer to [2], [25], [11], and for a more abstract development, we refer to the monograph [12]. These works showed that the cases where $g\le0$ (which corresponds to maximization of nonnegative rewards) and $g\ge0$ (which is most relevant to the control problems of this paper) are quite different in structure. In particular, while VI converges to $ J^*$ starting for $J_0\equiv0$ when $g\le0$, this is not so when $g\ge0$; a certain compactness condition is needed to guarantee this [see Example \ref{example2}, and part (d) of the following proposition]. Moreover when $g\ge0$, Bellman's equation may have solutions $\hat J\ne J^*$ with $\hat J\ge  J^*$ (see Example \ref{example1}), and VI will not converge to $ J^*$ starting from such $\hat J$. In addition it is known that in general, PI need not converge to $ J^*$ and may instead stop with a suboptimal policy  (see Example \ref{example3}). 

The following proposition gives the standard results when $g\ge0$ (see [2], Props.\ 5.2, 5.4, and 5.10, [11], Props.\ 4.1.1, 4.1.3, 4.1.5, 4.1.9, or [12], Props.\ 4.3.3, 4.3.9, and 4.3.14). These results hold for stochastic infinite horizon DP problems with nonnegative cost per stage, and do not take into account the favorable structure of deterministic problems or the presence of the stopping set $X_s$.

\begin{proposition}\label{propnegdp}
Let the nonnegativity condition (\ref{nonnegcost}) hold. 
\begin{itemize}
\item[]
\item[(a)] $ J^*$ satisfies Bellman's equation (\ref{bellmaneq}), and  if $\hat J\in E^+(X)$ is another solution, i.e., $\hat J$ satisfies
\begin{equation} \label{bellmaneqa}
\hat J(x)= \inf_{u\in U(x)}\big\{g(x,u)+\hat J\big(f(x,u)\big)\big\},\qquad \forall\ x\in X,
\end{equation}
 then  $ J^*\le \hat J$.
\item[(b)] For all stationary policies $\m$ we have 
\begin{equation} \label{polevala}
J_{\m}(x)=g\big(x,\m(x)\big)+J_{\m}\big(f\big(x,\m(x)\big)\big),\qquad \forall\ x\in X.
\end{equation}
\item[(c)] A stationary policy $\m^*$ is optimal if and only if 
\begin{equation} \label{polimprovea}
\m^*(x)\in \argmin_{u\in U(x)}\big\{g(x,u)+ J^*\big(f(x,u)\big)\big\},\qquad \forall\ x\in X.
\end{equation}
\item[(d)] If $U$ is a metric space and the sets 
\begin{equation} \label{onetwen}
U_k(x,\l) =\big\{ u\in U(x)\mid g(x,u)+J_k\big(f(x,u)\big)\le\l \big\}
\end{equation}
are compact for all $x\in X$, $\l\in \re$, and $k$, where $\{J_k\}$ is the sequence generated by  VI [cf.\ Eq.\ (\ref{vieq})] starting from $J_0\equiv0$, then there exists at least one optimal stationary policy, and we have $J_k\to J^*$.
\end{itemize}
\end{proposition}

Compactness assumptions such as the one of part (d) above, were originally given in [9], [10], and in [29]. They have been used in several other works, such as [3], [11], Prop.\ 4.1.9. In particular, the condition of part (d) holds when $U(x)$ is a finite set for all $x\in X$. The condition of part (d) also holds when $X=\rn$, and for each $x\in X$, the set
\begin{equation*}
\big\{u\in U(x)\mid g(x,u)\le\l\big\}
\end{equation*}
is a compact subset of $\re^m$, for all $\l\in\re$, and $g$ and $f$ are continuous in $u$. The proof consists of showing by induction that the VI iterates $J_k$ have compact level sets and hence are lower semicontinuous.

Let us also note a recent result of H.\ Yu and the author [35], where it was shown that $ J^*$ is the unique solution of Bellman's equation within the class of all functions $J\in E^+(X)$ that satisfy
\begin{equation} \label{bridgingcond}
0\le J\le c J^*\qquad\hbox{for some }c>0,
\end{equation}
(we refer to [35] for discussion and references to antecedents of this result). Moreover it was shown that VI converges to $ J^*$ starting from any function satisfying the  condition
$$ J^*\le J\le c J^*\qquad\hbox{for some }c>0,$$
and under the compactness conditions of Prop.\ \ref{propnegdp}(d), starting from any $J$ that satisfies Eq.\ (\ref{bridgingcond}). The same paper and a related paper [5] discuss extensively PI algorithms for stochastic nonnegative cost problems.

For deterministic problems, there has been substantial research in the adaptive dynamic programming literature, regarding the validity of Bellman's equation and the uniqueness of its solution, as well as the attendant questions of convergence of VI and PI. 
In particular, infinite horizon deterministic optimal control for both discrete-time and continuous-time systems has been considered since the early days of DP in the works of Bellman. For continuous-time problems the questions discussed in the present paper involve substantial technical difficulties, since the analog of the (discrete-time) Bellman equation (\ref{bellmaneq}) is the steady-state form of the (continuous-time) Hamilton-Jacobi-Bellman equation, a nonlinear partial differential equation the solution and analysis of which is in general very complicated. A formidable difficulty is the potential lack of differentiability of the optimal cost function, even for simple problems such as time-optimal control of second order linear systems to the origin. 

The analog of VI for continuous-time systems essentially involves the time integration of the Hamilton-Jacobi-Bellman equation, and its analysis must deal with difficult issues of stability and convergence to a steady-state solution. Nonetheless there have been proposals of continuous-time PI algorithms, in the early papers [26], [23], [28], [34], and the thesis [6], as well as more recently in several works; see e.g., the book [32], the survey [18], and the references quoted there. These works also address the possibility of value function approximation, similar to other approximation-oriented methodologies such as neurodynamic programming [4] and reinforcement learning [31], which consider primarily discrete-time systems. For example, among the restrictions of the PI method, is that it must be started with a stabilizing controller; see for example the paper [23], which considered linear-quadratic continuous-time problems, and showed convergence to the optimal policy of the PI algorithm, assuming that an initial stabilizing linear controller is used.
By contrast, no such restriction is needed in the PI methodology of the present paper; questions of stability are addressed only indirectly through the finiteness of the values $ J^*(x)$ and Assumption \ref{assumptiondetoc}.

For discrete-time systems there has been much research, both for VI and PI algorithms. For a  selective list of recent references, which themselves contain extensive lists of other references, see the book [32], the papers [19], [16], [17], [22], [33],  the survey
papers in the edited volumes [27] and [21], and the special issue  [20]. Some of these works relate to continuous-time problems as well, and in their treatment of algorithmic convergence, typically assume that $X$ and $U$ are Euclidean spaces, as well as continuity and other conditions on $g$, special structure of the system, etc. It is beyond our scope to provide a detailed survey of the state-of-the-art of the VI and PI methodology in the context of adaptive DP. However, it should be clear that the works in this field involve more restrictive assumptions than our corresponding results of Props.\ \ref{propdetoc1}-\ref{propdetpolicyit}. Of course, these  works also address questions that we do not, such as issues of stability of the obtained controllers, the use of approximations, etc.  Thus the results of the present work may be viewed as new in that they rely on very general assumptions, yet do not address some important practical issues. The line of analysis of the present paper, which is based on general results of Markovian decision problem theory and abstract forms of dynamic programming, is also different from the lines of analysis of works in adaptive DP, which make heavy use of the deterministic character of the problem and control theoretic methods such as Lyapunov stability.

Still there is a connection between our line of analysis and Lyapunov stability. In particular, if $\p^*$ is an optimal controller, i.e., $J_{\p^*}=J^*$, then for every $x_0\in X_f$, the state sequence $\{x_k\}$ generated using $\p^*$ and starting from $x_0$ remains within $X_f$ and satisfies $J^*(x_k)\downarrow 0$. This can be seen by writing
$$J^*(x_0)=\sum_{t=0}^{k-1}g\big(x_t,\m^*_t(x_t)\big)+J^*(x_{k}),\qquad k=1,2,\ldots,$$
and using the facts $g\ge0$ and $J^*(x_0)<\infty$. Thus an optimal controller, restricted to the subset $X_f$,  may be viewed as a Lyapunov-stable controller where the Lyapunov function is $J^*$. 

On the other hand, existence of a ``stable" controller does not necessarily imply that $J^*$ is real-valued. In particular, it may not be true that if the generated sequence $\{x_k\}$ by an optimal controller starting from some $x_0$ converges to $X_s$,  then we have $J^*(x_0)<\infty$. The reason is that the cost per stage $g$ may not decrease fast enough as we approach $X_s$. As an example, let 
$$X=\{0\}\cup\{1/m\mid m:\hbox{is a positive integer}\},$$
with $X_s=\{0\}$, and assume that there is a unique controller, which moves from $1/m$ to $1/(m+1)$ with incurred cost $1/m$. Then we have $J^*(x)=\infty$ for all $x\ne0$, despite the fact that the controller is ``stable" in the sense that it generates a sequence $\{x_k\}$ converging to 0 starting from every $x_0\ne0$.

\section{Proofs of the Main Results}

\pn Let us denote for all $x\in X$,
\begin{equation*}
\Pi_{T,x}=\big\{\p\in \Pi\mid \p\hbox{ terminates from }x\big\},
\end{equation*}
and note the following key implication of Assumption \ref{assumptiondetoc}: 
\begin{equation} \label{jstareq}
J^*(x)=\inf_{\p\in \Pi_{T,x}}J_\p(x),\qquad \forall\ x\in X_f.
\end{equation}
In the subsequent arguments, the significance of policies that terminate starting from some initial state $x_0$ is that the corresponding generated sequences $\{x_k\}$ satisfy $J(x_k)=0$ for all $J\in{\cal J}$ and $k$ sufficiently large.
\smskip

\pn {\bf Proof of Prop.\ \ref{propdetoc1}:}  Let $\hat J\in{\cal J}$ be a solution of the Bellman equation (\ref{bellmaneqa}), so that 
\begin{equation} \label{belineq}
\hat J(x)\le g(x,u)+\hat J\big(f(x,u)\big),\qquad \forall\ x\in X,\ u\in U(x),
\end{equation}
while by Prop.\ \ref{propnegdp}(a), $ J^*\le \hat J$. For any $x_0\in X_f$ and policy $\p=\{\m_0,\m_1,\ldots\}\in \Pi_{T,x_0}$, we have by using repeatedly Eq.\ (\ref{belineq}), 
\begin{equation*}
 J^*(x_0)\le \hat J(x_0)\le \hat J(x_k)+\sum_{t=0}^{k-1}g\big(x_t,\m_t(x_t)\big),\ \ k=1,2,\ldots,
\end{equation*}
where $\{x_k\}$ is the state sequence generated starting from $x_0$ and using $\p$.  Also, since $\p\in \Pi_{T,x_0}$ and hence $x_k\in X_s$ and $\hat J(x_k)=0$ for all sufficiently large $k$, we have 
\begin{align*}
\limsup_{k\to\infty}&\lf\{\hat J(x_k)+\sum_{t=0}^{k-1}g\big(x_t,\m_t(x_t)\big)\ri\}\\
&=\lim_{k\to\infty}\lf\{\sum_{t=0}^{k-1}g\big(x_t,\m_t(x_t)\big)\ri\}\\
&=J_\p(x_0).
\end{align*}
By combining the last two relations, we obtain
\begin{equation*}
J^*(x_0)\le \hat J(x_0)\le J_\p(x_0),\qquad \forall\ x_0\in X_f,\ \p\in \Pi_{T,x_0}.
\end{equation*}
Taking the infimum over $\p\in \Pi_{T,x_0}$ and using Eq.\  (\ref{jstareq}), it follows that $ J^*(x_0)=\hat J(x_0)$ for all $x_0\in X_f$. Also for $x_0\notin X_f$, we have $ J^*(x_0)=\hat J(x_0)=\infty$ [since $ J^*\le \hat J$ by Prop.\ \ref{propnegdp}(a)], so we obtain $ J^*=\hat J$. \qed

\smskip

 \pn {\bf Proof of Prop.\ \ref{propdetvi}:} (a) Suppose that $J_0\in {\cal J}$ and $J_0\ge J^*$. Starting with $J_0$, let us  apply the VI operation to both sides of the inequality $J_0\ge J^*$. Since $ J^*$ is a solution of Bellman's equation and VI has a monotonicity property that  maintains the direction of functional inequalities, we see that $J_1\ge  J^*$. Continuing similarly, we obtain $J_k\ge  J^*$ for all $k$. Moreover, we clearly have $J_k(x)=0$ for all $x\in X_s$, so $J_k\in{\cal J}$ for all $k$. We now argue that since $J_k$ is produced by $k$ steps of VI starting from $J_0$, it is the optimal cost function of the $k$-stage version of the problem with terminal cost function $J_0$. Therefore, we have for every $x_0\in X$ and policy $\p=\{\m_0,\m_1,\ldots\}$,
 \begin{equation*}
J^*(x_0)\le J_k(x_0)\le J_0(x_k)+\sum_{t=0}^{k-1}g\big(x_t,\m_t(x_t)\big),\ \ k=1,2,\ldots,
\end{equation*}
where $\{x_t\}$ is the state sequence generated starting from $x_0$ and using $\p$. If $x_0\in X_f$ and $\p\in \Pi_{T,x_0}$, we have $x_k\in X_s$ and $J_0(x_k)=0$ for all sufficiently large $k$, so that
\begin{align*}
\limsup_{k\to\infty}&\lf\{J_0(x_k)+\sum_{t=0}^{k-1}g\big(x_t,\m_t(x_t)\big)\ri\}\\
&=\lim_{k\to\infty}\lf\{\sum_{t=0}^{k-1}g\big(x_t,\m_t(x_t)\big)\ri\}\\
&=J_\p(x_0).
\end{align*}
By combining the last two relations, we obtain
\begin{equation*}
J^*(x_0)\le \liminf_{k\to\infty}J_k(x_0)\le \limsup_{k\to\infty}J_k(x_0)\le J_\p(x_0),
\end{equation*}
for all $x_0\in X_f$ and $\p\in \Pi_{T,x_0}$. Taking the infimum over $\p\in \Pi_{T,x_0}$ and using Eq.\ (\ref{jstareq}), it follows that $\lim_{k\to\infty}J_k(x_0)= J^*(x_0)$ for all $x_0\in X_f$.  Since for $x_0\notin X_f$, we have $ J^*(x_0)=J_k(x_0)=\infty$, we obtain $J_k\to J^*$. 
\smskip
\pn (b) Let $\{J_k\}$ be the VI sequence generated starting from some function $J\in{\cal J}$. By the monotonicity of the VI operation, $\{J_k\}$ lies between the sequence of VI iterates starting  from the zero function [which converges to $ J^*$ from below by Prop. \ref{propnegdp}(d)], and the sequence of VI iterates starting from $J_0=\max\{J, J^*\}$ [which converges to $ J^*$ from above by part (a)].  \qed

\smskip

\pn {\bf Proof of Prop.\ \ref{propdetpolicyit}:} If $\m$ is a stationary policy and $\bar \m$ satisfies the policy improvement equation 
\begin{equation*}
\bar \m(x)\in \argmin_{u\in U(x)}\big\{g(x,u)+J_{\m}\big(f(x,u)\big)\big\},\qquad x\in X,
\end{equation*}
[cf.\ Eq.\ (\ref{polimprove})], we have for all $x\in X$,
\begin{align}\label{polimprovement}
J_\mu(x)&=g\big(x,\m(x)\big)+J_{\m}\big(f\big(x,\m(x)\big)\big)\notag\\
&\ge \min_{u\in U(x)}\big\{g(x,u)+J_{\m}\big(f(x,u)\big)\big\}\\ 
&=g\big(x,\bar \m(x)\big)+J_{\m}\big(f\big(x,\bar \m(x)\big)\notag\big),
\end{align} 
where the first equality follows from Prop.\ \ref{propnegdp}(b) and the second equality follows from the definition of $\bar \m$. Let  us fix $x$ and let $\{x_k\}$ be the sequence generated starting from $x$ and using $\m$. By repeatedly applying  Eq.\ (\ref{polimprovement}), we see that the sequence $\big\{\tl J_{k}(x)\big\}$ defined by 
\begin{equation*}
\tl J_{0}(x)=J_\m(x),
\end{equation*}
\begin{equation*}
\tl J_{1}(x)=J_\m(x_1)+g\big(x,\bar \m(x)\big),
\end{equation*}
and more generally,
\begin{equation*}
\tl J_{k}(x)=J_\m(x_k)+\sum_{t=0}^{k-1}g\big(x_t,\bar \m(x_t)\big),\qquad k=1,2,\ldots,
\end{equation*}
is monotonically nonincreasing.
Thus, using also Eq.\ (\ref{polimprovement}), we have 
\begin{align*}
J_\m(x)&\ge \min_{u\in U(x)}\big\{g(x,u)+J_\m\big(f(x,u)\big)\big\}\\
&=\tl J_{1}(x)\\
&\ge\tl J_{k}(x),
\end{align*}
 for all $x\in X$ and $k\ge 1$. This implies that
\begin{align*}
J_\m(x)&\ge \min_{u\in U(x)}\big\{g(x,u)+J_\m\big(f(x,u)\big)\big\}\\
&\ge \lim_{k\to\infty}\tl J_k(x)\\
&=\lim_{k\to\infty}\lf\{J_\m(x_k)+\sum_{t=0}^{k-1}g\big(x_t,\bar \m(x_t)\big)\ri\}\\
&\ge \lim_{k\to\infty}\sum_{t=0}^{k-1}g\big(x_t,\bar \m(x_t)\big)\\
&=J_{\bar \m}(x),
\end{align*}
where the last inequality follows since $J_\m\ge0$. In conclusion, we have
\begin{equation*}
J_\m(x)\ge \inf_{u\in U(x)}\big\{g(x,u)+J_{\m}\big(f(x,u)\big)\big\}\ge J_{\bar \m}(x),\quad x\in X.
\end{equation*}

Using $\m^k$ and $\m^{k+1}$ in place of $\m$ and $\bar \m$ in the preceding relation, we obtain for all $x\in X$,
\begin{equation}\label{imprineqo}
J_{\m^{k}}(x)\ge \inf_{u\in U(x)}\big\{g(x,u)+J_{\m^k}\big(f(x,u)\big)\big\}\ge J_{\m^{k+1}}(x).
\end{equation}
Thus the sequence $\{J_{\m^k}\}$ generated by PI converges monotonically to some function $J_\infty\in E^+(X)$, i.e., $J_{\m^k}\downarrow J_\infty$. Moreover, by taking the limit as $k\to\infty$ in Eq.\ (\ref{imprineqo}), we have the two relations
\begin{equation*}
J_\infty(x)\ge \inf_{u\in U(x)}\big\{g(x,u)+J_\infty\big(f(x,u)\big)\big\},\qquad x\in X,
\end{equation*}
and
\begin{equation*}
g(x,u)+J_{\m^k}\big(f(x,u)\big)\ge J_\infty(x),\qquad x\in X,\ u\in U(x).
\end{equation*}
We now take the limit in the second relation as $k\to\infty$, then the infimum over $u\in U(x)$, and then combine with the first relation, to obtain
\begin{equation*}
J_\infty(x)= \inf_{u\in U(x)}\big\{g(x,u)+J_\infty\big(f(x,u)\big)\big\},\qquad x\in X.
\end{equation*}
Thus $J_\infty$ is a solution of Bellman's equation, satisfying $J_\infty\in{\cal J}$ (since $J_{\m^k}\in{\cal J}$ and $J_{\m^k}\downarrow J_\infty$), so  by the uniqueness result of Prop.\ \ref{propdetoc1}, we have $J_\infty= J^*$. \qed

\section{Discussion, Special Cases, and Extensions}

\pn 
In this section we elaborate on our main results and we derive easily verifiable conditions under which our assumptions hold. 

\subsection{Conditions that Imply Assumption \ref{assumptiondetoc}}

\pn Consider Assumption \ref{assumptiondetoc}. As noted in Section I, it  holds when $X$ and $U$ are finite, a terminating policy exists from every $x$, and all cycles of the state transition graph have positive length.
For the case where $X$ is infinite, let us assume that  $X$ is a normed space with norm denoted $\|\cdot\|$, and  say that $\pi$ {\it asymptotically terminates from $x$} if the sequence $\{x_k\}$ generated starting from  $x$ and using $\pi$ converges to $X_s$ in the sense that 
$$\lim_{k\to\infty}\hbox{dist}(x_k,X_s)=0,$$
where $\hbox{dist}(x,X_s)$ denotes the minimum distance from $x$ to $X_s$,
\begin{equation*}
\hbox{dist}(x,X_s)=\inf_{y\in X_s}\|x-y\|,\qquad x\in X.
\end{equation*}
The following proposition provides readily verifiable conditions that guarantee Assumption \ref{assumptiondetoc}.

\begin{proposition}\label{propdetassump}
Let the cost nonnegativity condition \eqref{nonnegcost} and stopping set conditions \eqref{absorbe}-\eqref{stopset} hold, and assume further 
the following:
\begin{itemize}
\item[]
\item[(1)] For every $x\in X_f$ and $\e>0$, there exits a policy $\p$ that asymptotically terminates from $x$ and satisfies 
\begin{equation*}
J_\p(x)\le  J^*(x)+\e.
\end{equation*}
\item[(2)] For every $\e>0$, there exists a $\d_\e>0$ such that for each $x\in  X_f$ with 
$$\hbox{dist}(x,X_s)\le \d_\e,$$
there is a policy $\pi$ that terminates from $x$ and satisfies $J_\p(x)\le \e$.
\end{itemize}
\smskip
\pn Then  Assumption \ref{assumptiondetoc}  holds.
\end{proposition}

\proof Fix $x\in X_f$ and $\e>0$.
Let $\p$ be a policy that asymptotically terminates from $x$, and satisfies $J_\pi(x)\le J^*(x)+\e$, as per condition (1). Starting from $x$, this policy will generate a sequence $\{x_k\}$ such that for some index $\bar k$ we have
\begin{equation*}
\hbox{dist}(x_{\bar k},X_s)\le \d_{\e},
\end{equation*}
so by condition (2), there exists a policy $\bar \p$ that terminates from $x_{\bar k}$ and is such that $J_{\bar \p}(x_{\bar k})\le \e$. Consider the policy $\p'$ that follows $\p$ up to index $\bar k$ and follows $\bar \p$ afterwards. This policy terminates from $x$ and satisfies
\begin{equation*}
J_{\p'}(x)=J_{\p,\bar k}(x)+J_{\bar \p}(x_{\bar k})\le J_{\p}(x)+J_{\bar \p}(x_{\bar k})\le  J^*(x)+2\e,
\end{equation*}
where $J_{\p,\bar k}(x)$ is the cost incurred by $\p$ starting from $x$ up to reaching $x_{\bar k}$. 
\qed

\smskip

Condition (1) of the preceding proposition requires that for states $x\in X_f$, the optimal cost $J^*(x)$ can be achieved arbitrarily  closely  with policies that asymptotically terminate from $x$. Problems for which condition (1) holds are those involving a cost per stage that is strictly positive outside of $X_s$. More precisely, condition (1) holds if for each $\d>0$ there exists $\e>0$ such that 
\begin{equation}\label{sufcondone}
\inf_{u\in U(x)}g(x,u)\ge\e,\qquad \forall\ x\in X\hbox{ such that } \hbox{dist}(x,X_s)\ge \d.
\end{equation}
Then for any $x$ and policy $\p$  that does not asymptotically terminate from $x$, we will have $J_\p(x)=\infty$, so that if $x\in X_f$, all policies $\p$ with $J_\p(x)<\infty$ must be asymptotically terminating from $x$. In applications, condition (1) is natural and consistent with the aim of steering the state towards the terminal set $X_s$ with finite cost. 
Condition (2)  is a ``controllability" condition implying that the state can be steered into $X_s$ with arbitrarily small cost from a starting state that is sufficiently close to $X_s$. 

\begin{example}[Linear System Case] \label{example4}Consider a linear system
\begin{equation*}
x_{k+1}=Ax_k+Bu_k,
\end{equation*}
where $A$ and $B$ are given matrices,
 with the terminal set being the origin, i.e., $X_s=\{0\}$. We assume the following:
\begin{itemize}
\item[(a)] $X=\rn$, $U=\re^m$, and there is an open sphere $R$ centered at the origin such that $U(x)$ contains $R$ for all $x\in X$.
\item[(b)] The system is controllable, i.e., one may drive the system from any state to the origin within at most $n$ steps using suitable controls, or equivalently that the matrix $[B\ AB\ \cdots A^{n-1}B]$ has rank $n$.
\item[(c)] $g$ satisfies
\begin{equation*}
0\le g(x,u)\le \beta\big(\|x\|^{p}+\|u\|^{p}\big),\qquad \forall\ (x,u)\in V,
\end{equation*}
where $V$ is some open sphere centered at the origin, $\beta,p$ are some positive scalars, and $\|\cdot\|$ is the standard Euclidean norm. 
\end{itemize}

\smskip

\pn Then condition (2) of Prop.\ \ref{propdetassump} is satisfied, while $x=0$ is cost-free and absorbing  [cf.\ Eq.\ \eqref{absorbe}]. Still, however, in the absence of additional assumptions, there may be multiple solutions to Bellman's equation within ${\cal J}$. 

As an example, consider the scalar  system $x_{k+1}=ax_k+u_k$ with $X=U(x)=\re$, and the quadratic cost $g(x,u)=u^2$. Then Bellman's equation has the form
\begin{equation*}
J(x)=\min_{u\in\re}\big\{u^2+J(ax+u)\big\},\qquad x\in \re,
\end{equation*}
and it is seen that the optimal cost function, $J^*(x)\equiv0$, is a solution. Let us assume that $a>1$ so the system is unstable (the instability of the system is important for the purpose of this example). Then it can be verified that the quadratic function $J(x)=(a^2-1)x^2$, which belongs to $\cal J$,  also solves Bellman's equation. This is a case where the algebraic Riccati equation associated with the problem has two nonnegative solutions because there is no cost on the state, and a standard observability condition for uniqueness of solution of the Riccati equation is violated.

If on the other hand,  in addition to (a)-(c), we assume that for some positive scalars $\g,p$, we have $\inf_{u\in U(x)}g(x,u)\ge \g\|x\|^{p}$ for all $x\in\rn$, then $J^*(x)>0$ for all $x\ne0$ [cf.\ Eq.\ (\ref{stopset})], while condition (1) of Prop.\ \ref{propdetassump} is satisfied as well [cf.\ Eq.\ (\ref{sufcondone})]. Then by Prop.\ \ref{propdetassump}, Assumption \ref{assumptiondetoc} holds, and Bellman's equation has a unique solution within $\cal J$.
\end{example}

There are straightforward extensions of the conditions of the  preceding example to a nonlinear system. Note that even for a controllable system, it is possible that there exist states from which the terminal set cannot be reached, because $U(x)$ may imply constraints on the magnitude of the control vector. Still the preceding analysis allows for this case.

\subsection{An Optimistic Form of PI}

\pn Let us consider a variant of PI where policies are evaluated inexactly, with a finite number of VIs. In particular,
this algorithm starts with some $J_0\in E(X)$, and generates a sequence of cost function and policy pairs $\{J_k,\m^k\}$ as follows: Given $J_k$, we generate $\m^k$ according to
\begin{equation} \label{optpione}
\m^k(x)\in\arg\min_{u\in U(x)}\big\{g(x,u)+J_k\big(f(x,u)\big)\big\},\qquad x\in X,
\end{equation}
and then we obtain $J_{k+1}$ with $m_k\ge 1$ VIs using $\m^k$:
\begin{equation} \label{optpitwo}
J_{k+1}(x_0)=J_k(x_{m_k})+\sum_{t=0}^{m_k-1}g\big(x_t,\m^k(x_t)\big),\qquad x_0\in X,
\end{equation}
where $\{x_t\}$ is the sequence generated  using $\m^k$ and starting from $x_0$, and $m_k$ are arbitrary positive integers. Here $J_0$ is a function in ${\cal J}$ that is required to satisfy 
\begin{equation} \label{optpicond}
J_0(x)\ge \inf_{u\in U(x)}\big\{g(x,u)+J_0\big(f(x,u)\big)\big\},\ \ \forall\ x\in X,\ u\in U(x).
\end{equation}
For example $J_0$ may be equal to the cost function of some stationary policy, or be the function that takes the value 0 for $x\in X_s$ and $\infty$ at $x\notin X_s$. Note that when $m_k\equiv 1$ the method is equivalent to VI, while the case $m_k=\infty$ corresponds to the standard PI considered earlier. In practice, the most effective value of $m_k$ may be found experimentally, with moderate values $m_k>1$ usually working best. We refer to the textbooks [25] and [11] for discussions of this type of inexact PI algorithm (in [25] it is called ``modified" PI, while in [11] it is called ``optimistic" PI).

 \begin{proposition}[Convergence of Optimistic PI] \label{propdetpolicyitopt}
Let Assumption \ref{assumptiondetoc} hold.
For the  PI algorithm (\ref{optpione})-(\ref{optpitwo}), where $J_0$ belongs to ${\cal J}$ and satisfies the condition (\ref{optpicond}), we have $J_{k}\downarrow  J^*$. 
\end{proposition}

\proof We have for all $x\in X$,
\begin{align*}
J_0(x)&\ge \inf_{u\in U(x)}\big\{g(x,u)+J_0\big(f(x,u)\big)\big\}\\
&=g\big(x,\m^0(x)\big)+J_0\big(f(x,\m^0(x))\big)\\
&\ge J_1(x)\\
&\ge g\big(x,\m^0(x)\big)+J_1\big(f(x,\m^0(x))\big)\\
&\ge \inf_{u\in U(x)}\big\{g(x,u)+J_1\big(f(x,u)\big)\big\}\\
&= g\big(x,\m^1(x)\big)+J_1\big(f(x,\m^1(x))\big)\\
&\ge J_2(x),
\end{align*}
where the first inequality is the condition (\ref{optpicond}), the second and third inequalities follow because of the monotonicity of the $m_0$ value iterations (\ref{optpitwo}) for $\m^0$, and the fourth inequality follows from the policy improvement equation (\ref{optpione}). Continuing similarly, we have 
\begin{equation*}
J_k(x)\ge \inf_{u\in U(x)}\big\{g(x,u)+J_k\big(f(x,u)\big)\big\}\ge J_{k+1}(x),
\end{equation*}
for all $x\in X$ and $k$. Moreover, since $J_0\in{\cal J}$, we have $J_k\in{\cal J}$ for all $k$. Thus  $J_{k}\downarrow J_\infty$ for some $J_\infty\in {\cal J}$, and similar to the proof of Prop.\ \ref{propdetpolicyit}, it follows that $J_\infty$ is a solution of Bellman's equation. Hence,  by the uniqueness result of Prop.\ \ref{propdetoc1}, we have $J_\infty= J^*$. \qed

\subsection{Minimax Control to a Terminal Set of States}

\pn Our analysis can be readily extended to minimax problems with a terminal set of states. Here the system is
\begin{equation*}
x_{k+1}=f(x_k,u_k,w_k),\qquad k=0,1,\ldots,
\end{equation*}
where $w_k$ is the control of an antagonistic opponent that aims to maximize the cost function. We assume that $w_k$ is chosen from a given set $W$ to maximize the sum of costs per stage, which are assumed nonnegative:
\begin{equation*}
0\le g(x,u,w)\le\infty,\qquad x\in X,\ U\in U(x),\ w\in W.
\end{equation*}

We wish to choose a policy $\pi=\{\m_0,\m_1,\ldots\}$ to minimize the cost function
\begin{equation*}
J_\p(x_0)=\sup_{w_k\in W\atop k=0,1,\ldots}\lim_{k\to\infty}\sum_{t=0}^k g\big(x_k,\m_k(x_k),w_k\big),
\end{equation*}
where $\big\{x_k,\m_k(x_k)\big\}$ is a state-control sequence corresponding to $\p$ and the sequence $\{w_0,w_1,\ldots\}$.
We assume that there is a termination set $X_s$, the states of which are cost-free and absorbing,  i.e., 
\begin{equation*}
g(x,u,w)=0,\qquad x=f(x,u,w),
\end{equation*}
for all $x\in X_s$, $u\in U(x)$, $w\in W$, and that all states outside $X_s$ have strictly positive optimal cost, so that 
\begin{equation*}
X_s=\big\{x\in X\mid  J^*(x)=0\big\}.
\end{equation*}
The finite-state version of this problem has been discussed in 
[13], under the name {\it robust shortest path planning\/}, for the case where $g$ can take both positive and negative values. A  problem that is closely related is {\it reachability of a target set in minimum time\/}, which is obtained for 
\begin{equation*}
g(x,u,w)=\begin{cases}
0& \text{if $x\in X_s$,}\\
1& \text{if $x\notin X_s$,}
\end{cases}
\end{equation*}
assuming also that the control process stops once the state enters the set $X_s$. Here $w$ is a disturbance described by set membership ($w\in W$), and the objective is to reach the set $X_s$ in the minimum guaranteed number of steps. The set $X_f$ is the set of states for which $X_s$ is guaranteed to be reached in a finite number of steps. Another related problem is {\it reachability of a target tube\/}, where for a given set $\hat X$, 
\begin{equation*}
g(x,u,w)=\begin{cases}
0& \text{if $x\in \hat X$,}\\
1& \text{if $x\notin \hat X$,}
\end{cases}
\end{equation*}
and the objective is to find the initial states starting from which we can guarantee to keep all future states within $\hat X$. These  two reachability problems  were first formulated and analyzed as part of the author's Ph.D.\ thesis research [7], and the subsequent paper [8]. In fact the reachability algorithms given in these works are essentially special cases of the VI algorithm of the present paper, starting with appropriate initial functions $J_0$.

To extend our results to the general form of the minimax problem described above, we need to adapt the definition of termination. In particular, given a state $x$, in the minimax context we say that a policy $\pi$ terminates from $x$ if there exists an index $\bar k$ [which depends on $(\p,x)$] such that the sequence $\{x_k\}$, which is generated starting from $x$ and using $\pi$, satisfies $x_{\bar k}\in X_s$ for all sequences $\{w_0,\ldots,w_{\bar k-1}\}$ with $w_t\in W$ for all $t=0,\ldots\bar k-1$. Then Assumption \ref{assumptiondetoc} is modified to reflect this new definition of termination, and our results can be readily extended, with Props.\ \ref{propdetoc1}, \ref{propdetvi}, \ref{propdetpolicyit}, and \ref{propdetpolicyitopt},  and their proofs, holding essentially as stated. The main adjustment needed is to replace expressions of the forms
\begin{equation*}
g(x,u)+J\big(f(x,u)\big)
\end{equation*}
and
\begin{equation*}
J(x_k)+\sum_{t=0}^{k-1}g(x_t,u_t)
\end{equation*}
in these proofs with
\begin{equation*}
\sup_{w\in W}\big\{g(x,u,w)+J\big(f(x,u,w)\big)\big\}
\end{equation*}
and
\begin{equation*}
\sup_{w_t\in W\atop t=0,\ldots,k-1}\lf\{J(x_k)+\sum_{t=0}^{k-1}g(x_t,u_t,w_t)\ri\},
\end{equation*}
respectively; see also [14] for a more abstract view of such lines of argument.

\vskip-1pc

\section{Concluding Remarks}

\pn In this paper we have considered problems of deterministic optimal control to a terminal set of states subject to very general assumptions. Under reasonably practical conditions, we have established the uniqueness of solution of Bellman's equation, and the convergence of value and policy iteration algorithms,  even when there are states with infinite optimal cost. Our analysis bypasses the need for assumptions involving the existence of globally stabilizing controllers, which guarantee that the optimal cost function $ J^*$ is real-valued. This generality makes our results a convenient starting point for analysis of problems involving additional assumptions, and perhaps cost function approximations. 

While we have restricted attention to undiscounted problems, the line of analysis of the present paper applies also to discounted problems with one-stage cost function $g$ that may be unbounded from above. Similar but more favorable results can be obtained, thanks to the presence of the discount factor; see the author's paper [14], which contains related analysis for stochastic and minimax, discounted and undiscounted problems, with nonnegative cost per stage. 

The  results for these problems, and the results of the present paper, have a common ancestry. They fundamentally draw their validity from notions of regularity, which were developed in the author's abstract DP monograph [12] and were extended recently in [14]. Let us describe the regularity idea briefly, and its connection to the analysis of this paper. Given a set of functions $S\in E^+(X)$, we say that a collection ${\cal C}$ of policy-state pairs $(\p,x_0)$, with $\p\in \Pi$ and $x_0\in X$,  is {\it $S$-regular}  if for all $(\p,x_0)\in {\cal C}$ and  $J\in S$, we have
\begin{equation*}
J_\p(x_0)= \lim_{k\to\infty}\lf\{J(x_k)+ \sum_{t=0}^{k-1}
g\bl(x_t,\mu_t(x_t)\br)\ri\}.
\end{equation*}
In words, for all $(\p,x_0)\in {\cal C}$,
$J_\p(x_0)$ can be obtained in the limit by VI starting from any $J\in S$. The favorable properties with respect to VI of an $S$-regular collection  ${\cal C}$ can be translated into interesting properties relating to solutions of Bellman's equation and convergence of VI. In particular, the optimal cost function over the set of policies $\{\p\,\mid\, (\p,x)\in{\cal C}\}$,
$$J^*_{\cal C}(x)=\inf_{ \{\p\,\mid\, (\p,x)\in{\cal C}\}}J_\p(x),\qquad x\in X,$$
under appropriate problem-dependent assumptions, is the unique solution of Bellman's equation within the set $\big\{J\in S\mid J\ge J^*_{\cal C}\big\}$, and can be obtained by VI starting from any $J$ within that set (see [14]).

Within the deterministic optimal control context of this paper, it works well to choose ${\cal C}$ to be the set of all $(\p,x)$ such that $x\in X_f$ and $\p$ is terminating starting from $x$, and to choose $S$ to be ${\cal J}$, as defined by Eq.\ (\ref{regsets}). Then, in view of Assumption 1, we have $J^*_{\cal C}=J^*$, and the favorable properties of $J^*_{\cal C}$ are shared by $J^*$. For other types of problems
 different choices of  ${\cal C}$ may be appropriate, and corresponding results relating to the uniqueness of solutions of Bellman's equation and the validity of value and policy iteration may be obtained; see [14].


\begin{IEEEbiography}{Dimitri P.\ Bertsekas} studied engineering at the National Technical University of Athens, Greece, obtained his MS in electrical engineering at the George Washington University, Wash.\ DC in 1969, and his Ph.D.\ in system science in 1971 at the Massachusetts Institute of Technology (M.I.T.).

Dr. Bertsekas has held faculty positions with the Engineering-Economic Systems Dept., Stanford University (1971-1974) and the Electrical Engineering Dept.\ of the University of Illinois, Urbana (1974-1979). Since 1979 he has been teaching at the Electrical Engineering and Computer Science Department of M.I.T., where he is currently McAfee Professor of Engineering. He consults regularly with private industry and has held editorial positions in several journals. His research has spanned several fields, including optimization, control, large-scale and distributed computation, and data communication networks, and is closely tied to his teaching and book authoring activities. He has written numerous research papers, and sixteen books, several of which are used as textbooks in M.I.T.\ classes.

Professor Bertsekas was awarded the INFORMS 1997 Prize for Research Excellence in the Interface Between Operations Research and Computer Science for his book ``Neuro-Dynamic Programming" (co-authored with John Tsitsiklis), the 2001 ACC John R.\ Ragazzini Education Award, the 2009 INFORMS Expository Writing Award,  the 2014 ACC Richard E.\ Bellman Control Heritage Award, the 2014 Khachiyan Prize for Life-Time Accomplishments in Optimization, and the SIAM/MOS 2015 George B.\ Dantzig Prize. In 2001 he was elected to the United States National Academy of Engineering.

Dr.\ Bertsekas' recent books are ``Dynamic Programming and Optimal Control: 4th Edition" (2012), ``Abstract Dynamic Programming" (2013), and ``Convex Optimization Algorithms" (2015), all published by Athena Scientific. 
\end{IEEEbiography}

\bibliography{TD_Methods_ref}
\bibliographystyle{IEEEtran}

\section{References}
\def\ref{\vskip1.pt\pn}

\def\refer{\ref}

\ref[1] Bertsekas, D.\ P., and Rhodes, I.\ B., 1971.\ ``On the Minimax Reachability of Target Sets and Target Tubes," Automatica, Vol. 7, pp. 233-241.

\ref [2]  Bertsekas, D.\ P., and Shreve, S.\ E., 1978.\  Sto\-chastic Optimal
Control:  The Discrete Time Case, Academic Press, N.\ Y.; may be downloaded 
from http://web.mit.edu/dimitrib/www/home.html

\ref [3]  Bertsekas, D.\ P., and Tsitsiklis, J.\ N., 1991.\ ``An Analysis of
Stochastic Shortest Path Problems,"
Math.\ of Operations Research, Vol.\ 16, pp.\ 580-595.

\ref [4]  Bertsekas, D.\ P., and Tsitsiklis, J.\ N., 1996.\ Neuro-Dynamic
Programming, Athena Scientific, Belmont, MA.

\ref[5] Bertsekas, D.\ P., and Yu, H., 2015.\ ``Stochastic Shortest Path Problems Under Weak Conditions,"  Lab.\ for Information and Decision Systems Report LIDS-2909, revision of March 2015, to appear in Math.\ of Operations Research.

\ref[6] Beard, R.\ W., 1995.\ Improving the Closed-Loop Performance of Nonlinear Systems, Doctoral dissertation, Rensselaer Polytechnic Institute.

\ref [7] Bertsekas, D.\ P., 1971.\ ``Control of Uncertain Systems With a
Set-Member\-ship Description of the Uncertainty," Ph.D.\ Thesis, Dept.\ of EECS, 
MIT; may be downloaded  from\hfill\break  http://web.mit.edu/dimitrib/www/publ.html.

\ref [8] Bertsekas, D.\ P., 1972.\  ``Infinite Time Reachability of State
Space Regions by Using Feedback Control," IEEE Trans.\ Automatic Control, Vol.
AC-17, pp.\ 604-613.

\ref [9] Bertsekas, D.\ P., 1975.\  ``Monotone Mappings in Dynamic Programming," Proc.\ 1975 IEEE Conference on Decision and Control, Houston, TX, pp.\ 20-25. 

\ref [10] Bertsekas, D.\ P., 1977.\  ``Monotone Mappings with Application in
Dynamic Programming," SIAM J.\ on Control and Optimization, Vol.\ 15, pp.\
438-464.

\ref[11] Bertsekas, D.\ P., 2012.\ Dynamic Programming and Optimal Control, Vol.\ II: Approximate Dynamic Programming, Athena Scientific, Belmont, MA.

\ref[12] Bertsekas, D.\ P., 2013.\ Abstract Dynamic Programming, Athena Scientific, Belmont, MA.

\ref[13] Bertsekas, D.\ P., 2014.\ ``Robust Shortest Path Planning and Semicontractive Dynamic Programming,"  Lab.\ for Information and Decision Systems Report LIDS-P-2915, Feb.\ 2014 (revised Jan. 2015). 

\ref[14] Bertsekas, D.\ P., 2015.\ ``Regular Policies in Abstract Dynamic Programming,"  Lab.\ for Information and Decision Systems Report LIDS-3173, April 2015.

\ref [15] Blackwell, D., 1965.\  ``Positive Dynamic Programming," Proc.\ Fifth Berkeley Symposium Math.\ Statistics and Probability, pp.\ 415-418.

\ref[16] Heydari, A., 2014.\ ``Revisiting Approximate Dynamic Programming and its Convergence," IEEE Transactions on Cybernetics, Vol.\ 44, pp.\ 2733-2743.

\ref[17] Heydari, A., 2014.\ ``Stabilizing Value Iteration With and Without Approximation Errors," available at arXiv:1412.5675.

\ref[18] Jiang, Y., and Jiang, Z.\ P., 2013.\ ``Robust Adaptive Dynamic Programming for Linear and Nonlinear Systems: An Overview," Eur.\ J.\ Control, Vol.\ 19, pp.\ 417-425.

\ref[19] Jiang, Y., and Jiang, Z.\ P., 2014.\ ``Robust Adaptive Dynamic Programming and Feedback Stabilization of Nonlinear Systems," IEEE Trans.\ on Neural Networks and Learning Systems, Vol.\ 25, pp.\ 882-893.

\ref[20] Lewis, F.\ L., Liu, D., and Lendaris, G.\ G., 2008.\ Special Issue on Adaptive Dynamic Programming and Reinforcement Learning in Feedback Control, IEEE Trans.\ on Systems, Man, and Cybernetics, Part B, Vol.\ 38, No.\ 4.

\ref[21] Lewis, F.\ L., and Liu, D., (Eds), 2013.\ Reinforcement Learning and Approximate Dynamic Programming for Feedback Control, Wiley, Hoboken, N.\ J.

\ref[22] Liu, D., and Wei, Q., 2013.\ ``Finite-Approximation-Error-Based Optimal Control Approach for Discrete-Time Nonlinear Systems, IEEE Transactions on Cybernetics, Vol.\ 43, pp.\ 779-789.

\ref [23] Kleinman, D.\ L., 1968.\  ``On an Iterative Technique for Riccati
Equation Computations," IEEE Trans.\ Automatic Control, Vol.\ AC-13, pp.\ 114-115.

\ref[24] Pallu de la Barriere, R., 1967.\ Optimal Control Theory, Saunders, Phila; reprinted by Dover, N. Y., 1980.

\ref [25] Puterman, M.\ L., 1994.\  Markov Decision Processes: Discrete Stochastic Dynamic Programming, J.\ Wiley, N.\ Y.

\ref[26] Rekasius, Z.\ V., 1964.\ ``Suboptimal Design of Intentionally Nonlinear Controllers," IEEE Trans.\ on Automatic Control, Vol.\ 9, pp.\ 380-386.

\ref[27] Si, J., Barto, A., Powell, W., and Wunsch, D., (Eds.) 2004.\ Learning and Approximate Dynamic
Programming, IEEE Press, N.\ Y.

\ref[28] Saridis, G.\ N., and Lee, C.-S.\ G., 1979.\ ``An Approximation Theory of Optimal Control for Trainable Manipulators," IEEE Trans.\ Syst., Man, Cybern., Vol. 9, pp.\ 152-159.

\ref [29] Schal, M., 1975.\  ``Conditions for Optimality in Dynamic
Programming and for the Limit of $n$-Stage Optimal Policies to be Optimal," Z.\
Wahrscheinlichkeitstheorie und Verw.\ Gebiete, Vol.\ 32, pp.\ 179-196.

\ref [30] Strauch, R., 1966.\  ``Negative Dynamic Programming," Ann.\ Math.\
Statist., Vol.\ 37, pp.\ 871-890.

\ref[31] Sutton, R.\  S., and Barto, A.\ G., 1998.\ Reinforcement Learning, MIT
Press, Cambridge, MA.

\ref[32] Vrabie, D., Vamvoudakis, K.\ G., and Lewis, F.\ L., 2013.\ Optimal Adaptive Control and Differential Games by Reinforcement Learning Principles,
The Institution of Engineering and Technology, London.

\ref[33] Wei, Q., Wang, F.\ Y., Liu, D., and Yang, X., 2014.\ ``Finite-Approximation-Error-Based Discrete-Time Iterative Adaptive Dynamic Programming," IEEE Transactions on Cybernetics, Vol.\ 44, pp.\ 2820-2833.

\ref[34] Werbos, P.\ J., 1992.\ ``Approximate Dynamic Programming for Real-Time Control and Neural Modeling," in Handbook of Intelligent Control (D.\ A.\ White and D.\ A.\ Sofge, eds.), Multiscience Press.

\ref[35]  Yu, H.,  and Bertsekas, D.\ P., 2013.\ ``A Mixed Value and Policy Iteration Method for Stochastic Control with Universally Measurable Policies," Lab.\ for Information and Decision Systems Report LIDS-P-2905, MIT; to appear in Math.\ of Operations Research.

\end{document}